\documentclass[letterpaper,english,aps,preprint,prb,endfloats]{revtex4}
\usepackage{mathptmx}
\usepackage[T1]{fontenc}
\usepackage[latin1]{inputenc}
\usepackage{graphicx}
\usepackage{amssymb}

\makeatletter

\makeatletter
\def\@dotsep{4.5}
\makeatother

\usepackage{babel}
\makeatother

\begin{document}

\title{Quantum-Classical Liouville Dynamics in the Mapping Basis }

\author{Hyojoon Kim}

\email{hkim@donga.ac.kr}

\affiliation{Department of Chemistry, Dong-A University, Hadan-2-dong, Busan 604-714,
Korea}

\author{Ali Nassimi}

\email{anassimi@chem.utoronto.ca}

\author{Raymond Kapral}

\email{rkapral@chem.utoronto.ca}

\affiliation{Chemical Physics Theory Group, Department of Chemistry, University
of Toronto, Toronto, Ontario M5S 3H6, Canada}

\begin{abstract}
The quantum-classical Liouville equation describes the dynamics of
a quantum subsystem coupled to a classical environment. It has been
simulated using various methods, notably, surface-hopping schemes.
A representation of this equation in the mapping Hamiltonian basis
for the quantum subsystem is derived. The resulting equation of motion,
in conjunction with expressions for quantum expectation values in
the mapping basis, provide another route to the computation of the
nonadiabatic dynamics of observables that does not involve surface-hopping
dynamics. The quantum-classical Liouville equation is exact for the
spin-boson system. This well-known model is simulated using an approximation
to the evolution equation in the mapping basis and close agreement
with exact quantum results is found. 
\end{abstract}
\maketitle

\section{Introduction}

Nonadiabatic quantum mechanical effects are known to be important
for the description of the dynamics of many chemical and biological
processes. Photochemical dynamics, proton and electron transfer reactions
and vibrational relaxation processes are just a few examples where
quantum effects play significant roles. Due to the difficulty of simulating
the full quantum dynamics of large, complex, many-body systems, various
mixed quantum-classical and semiclassical schemes have been developed.
Here we consider quantum dynamics based on the quantum-classical Liouville
equation (see Ref.~{[}\onlinecite{kapral06}] and references therein),
\begin{eqnarray}
\frac{d}{dt}\hat{\rho}_{W}(R,P,t) & = & -\frac{i}{\hbar}[\hat{H}_{W},\hat{\rho}_{W}(t)]\label{eq:QCrho}\\
 &  & +\frac{1}{2}(\{\hat{H}_{W},\hat{\rho}_{W}(t)\}-\{\hat{\rho}_{W}(t),\hat{H}_{W}\}),\nonumber \end{eqnarray}
 where $[\hat{A},\hat{B}]$ is the commutator and $\{\hat{A},\hat{B}\}$
is the Poisson bracket for any operators $\hat{A}$ and $\hat{B}$.
The density matrix $\hat{\rho}_{W}(R,P,t)$ is a function of the environmental
phase space variables $(R,P)$ and is an operator in the degrees of
freedom of the quantum subsystem. The Hamiltonian $\hat{H}_{W}$ includes
terms describing the quantum subsystem, its environment and the coupling
between these parts of the system. This equation has been used to
the describe nonadiabatic dynamics on coupled electronic states \citep{martens97,donoso98},
vibrational dephasing \citep{riga06}, proton transfer reactions \citep{hanna05,hanna06,kim06a,kim06e,kim08}
and population relaxation in the spin-boson model \citep{mackernan02,mackernan08},
to name a few examples. The simulation of the dynamics using this
equation presents challenges and a number of different schemes have
been devised for this purpose. Often the simulation methods are based
on specific representations of the quantum degrees of freedom. For
example, surface-hopping dynamics that make use of the adiabatic basis
have been constructed\citep{santer01,mackernan02,mackernan08}, evolution
of the density matrix in the diabatic basis has been carried out using
a trajectory-based algorithm\citep{donoso98} and a representation
of the dynamics in the force basis has been simulated using the multithreads
algorithm\citep{wan00,wan02}.

The discrete quantum degrees of freedom of the system can be described
by the {}``classical electron analog'' model\citep{miller01} or
the mapping formalism\citep{meyer79,stock95,stock97,muller98}. Extending
Schwinger's angular momentum formalism\citep{0chap-schwinger65} to
the $N$-level case, the mapping formulation employs a quantum-mechanically
exact mapping of discrete electronic states onto continuous variables;
thus, the dynamics of both electronic and nuclear degrees of freedom
are described by continuous variables~\citep{thoss99}. The mapping
basis has been used to compute quantum dynamics in the context of
semiclassical path integral formulations of the theory\citep{stock95,stock97,sun98,miller01}
and in linearized path integral methods\citep{bonella03,bonella05,bonella052}.
In this paper we show how the quantum-classical Liouville equation
can be written in this mapping representation. The resulting evolution
equation, like the basis-free quantum-classical Liouville equation
(\ref{eq:QCrho}) from which it was derived, provides a useful description
of the dynamics of a quantum subsystem coupled to its environment.
Since the quantum-classical Liouville equation is exact for any quantum
system bilinearly coupled to a harmonic bath, so is its representation
in the mapping basis presented here. The spin-boson model is of this
type and this standard test model, for which exact quantum results
are available, is employed to illustrate features of the simulation
of the mapping form of the quantum-classical Liouville equation. In
particular, we show that an approximation to the evolution operator
allows one to accurately simulate the evolution using few trajectories
with an algorithm that does not involve surface hopping dynamics.
Comparisons with the results of other simulation algorithms are made.
A discussion of the applicability of this representation of the theory
to general many-body quantum systems is given in the last section
of the paper.

\section{Quantum-classical dynamics in the mapping basis}

We consider a quantum mechanical system that is partitioned into a
subsystem and bath. The expectation value of an operator $\hat{B}(t)$
can be written generally as \begin{equation}
\overline{B(t)}={\rm {Tr}}\;\hat{B}(t)\hat{\rho}={\rm {Tr'}}\int dX\;\hat{B}_{W}(X,t)\hat{\rho}_{W}(X),\label{average}\end{equation}
 where a partial Wigner transform over bath degrees of freedom, \begin{equation}
\hat{B}_{W}(X)=\int dZe^{iP\cdot Z/\hbar}\langle R-\frac{Z}{2}|\hat{B}|R+\frac{Z}{2}\rangle,\end{equation}
 has been taken. Here, $X=(R,P)$ denotes phase space variables of
the bath. The initial density matrix is $\hat{\rho}_{W}(X)$. The
partially Wigner transformed Hamiltonian of the system can be written
as \begin{equation}
\hat{H}_{W}=\frac{{P}^{2}}{2M}+\frac{\hat{p}^{2}}{2m}+\hat{V}_{s}(\hat{q})+V_{B}({R})+\hat{V}_{c}(\hat{q},{R}),\label{Hamiltonian}\end{equation}
 where the subscripts $s$, $B$ and $c$ denote the subsystem, bath
and coupling, respectively. Letting $\hat{h}_{s}=\hat{p}^{2}/2m+\hat{V}_{s}(\hat{q})$
be the subsystem Hamiltonian, whose eigenvalue problem is $\hat{h}_{s}|\lambda\rangle=\epsilon_{\lambda}|\lambda\rangle$,
we can write the expectation value of $\hat{B}(t)$ in the form \begin{equation}
\overline{B(t)}=\sum_{\lambda,\lambda'}\int dX\; B_{W}^{\lambda\lambda'}(X,t)\rho_{W}^{\lambda'\lambda}(X).\label{average-sub}\end{equation}
 in the subsystem basis.

\subsection{Average value in mapping basis}

Next, we write this expectation value in the mapping basis by noting
that any operator $\hat{B}_{W}(X)$ can be decomposed as $\hat{B}_{W}(X)=\sum_{\lambda\lambda'}B_{W}^{\lambda\lambda'}(X)|\lambda\rangle\langle\lambda'|$.
The evolution of the $N$-state subsystem can be conveniently replaced,
using mapping relations, with that of $N$ fictitious harmonic oscillators
with occupation numbers limited to 0 or 1, namely, $|\lambda\rangle\rightarrow|m_{\lambda}\rangle=|0_{1},\cdots,1_{\lambda},\cdots0_{n}\rangle$
\citep{miller01,meyer79,stock95,stock97,muller98,thoss99,sun98,bonella03,bonella05,bonella052}.
The matrix element of an operator may then be written in the mapping
form, $B_{W}^{\lambda\lambda'}(X)=\langle\lambda|\hat{B}_{W}(X)|\lambda'\rangle=\langle m_{\lambda}|\hat{B}_{m}(X)|m_{\lambda'}\rangle$,
where \begin{equation}
\hat{B}_{m}(X)=\sum_{\lambda\lambda'}B_{W}^{\lambda\lambda'}(X)\hat{a}_{\lambda}^{\dag}\hat{a}_{\lambda'}.\end{equation}
 The mapping annihilation and creation operators are given by \begin{equation}
\hat{a}_{\lambda}=\sqrt{\frac{1}{2\hbar}}(\hat{q}_{\lambda}+i\hat{p}_{\lambda}),\quad\hat{a}_{\lambda}^{\dag}=\sqrt{\frac{1}{2\hbar}}(\hat{q}_{\lambda}-i\hat{p}_{\lambda}),\label{creation}\end{equation}
 and satisfy the commutation relation $[\hat{a}_{\lambda},\hat{a}_{\lambda'}^{\dag}]=\delta_{\lambda\lambda'}$.
Explicitly, we may write \begin{equation}
\hat{a}_{\lambda}^{\dag}\hat{a}_{\lambda'}=\frac{1}{2\hbar}[\hat{q}_{\lambda}\hat{q}_{\lambda'}+\hat{p}_{\lambda}\hat{p}_{\lambda'}-i(\hat{p}_{\lambda}\hat{q}_{\lambda'}-\hat{q}_{\lambda}\hat{p}_{\lambda'})].\label{ceation_annialation_phase_space}\end{equation}
 One may easily verify that the matrix elements of $\hat{B}_{m}(X)$
in the mapping basis are identical to those of $\hat{B}_{W}(X)$ in
the subsystem basis.

In the analysis that follows it is convenient to work in a Wigner
representation of the mapping basis. To this end we introduce a coordinate
representation of the mapping states and annihilation and creation
operators, \begin{eqnarray}
\langle q|m_{\lambda}\rangle & = & \langle q_{1},q_{2},...,q_{N}|0_{1},...,1_{\lambda},...,0_{N}\rangle\nonumber \\
 & = & \phi_{0}(q_{1})...\phi_{0}(q_{\lambda-1})\phi_{1}(q_{\lambda})...\phi_{0}(q_{N})\label{coordinatereporesentationofmapping}\end{eqnarray}
 and \begin{equation}
\langle q|\hat{a}_{\lambda}|q'\rangle=\frac{1}{\sqrt{2\hbar}}(q'_{\lambda}+\hbar\frac{\partial}{\partial q'_{\lambda}})\delta(q_{\lambda}-q'_{\lambda})\prod_{\mu\ne\lambda}^{N}\delta(q_{\mu}-q'_{\mu}),\label{coordinaterepresentationofannialation}\end{equation}
 with an analogous expression for $\langle q|\hat{a}_{\lambda}^{\dag}|q'\rangle$.
Here \begin{eqnarray}
\phi_{0}(q_{\lambda}) & = & (\pi\hbar)^{-1/4}e^{-q_{\lambda}^{2}/2\hbar},\nonumber \\
\phi_{1}(q_{\lambda}) & = & \sqrt{2}(\pi\hbar^{3})^{-1/4}q_{\lambda}e^{-q_{\lambda}^{2}/2\hbar}.\label{oscillator}\end{eqnarray}

Equation~(\ref{average-sub}) may be written in the mapping basis
using the coordinate representation to obtain \begin{eqnarray}
 &  & \overline{B(t)}=\sum_{\lambda,\lambda'}\int dX\langle m_{\lambda}|\hat{B}_{m}(X,t)|m_{\lambda'}\rangle\nonumber \\
 &  & \qquad\qquad\qquad\times\langle m_{\lambda'}|\hat{\rho}_{m}(X)|m_{\lambda}\rangle\nonumber \\
 &  & \quad=\sum_{\lambda,\lambda'}\int dX\int dqdq'dq''dq'''\langle m_{\lambda}|q\rangle\langle q|\hat{B}_{m}(X,t)|q'\rangle\nonumber \\
 &  & \qquad\langle q'|m_{\lambda'}\rangle\langle m_{\lambda'}|q''\rangle\langle q''|\hat{\rho}_{m}(X)|q'''\rangle\langle q'''|m_{\lambda}\rangle.\label{average-subm}\end{eqnarray}
 Note that the coordinate space dimension of the mapping variables
is $N$. We may now introducing the Wigner transforms of the coordinate
space matrix elements of the mapping variables, \begin{eqnarray}
 &  & \langle r-\frac{z}{2}|\hat{B}_{m}(X,t)|r+\frac{z}{2}\rangle=\frac{1}{(2\pi\hbar)^{N}}\int dp\; e^{-ipz/\hbar}B_{m}(x,X,t)\nonumber \\
 &  & \langle r'-\frac{z'}{2}|\hat{\rho}_{m}(X)|r'+\frac{z'}{2}\rangle=\int dp'e^{-ip'z'/\hbar}\rho_{m}(x',X),\end{eqnarray}
 where $x=(r,p)$ are the phase space coordinates of the mapping variables.
Using these definitions Eq.~(\ref{average-subm}) can be written
as \begin{eqnarray}
\overline{B(t)} & = & \int dXdxdx'\; B_{m}(x,X,t)f(x,x')\rho_{m}(x',X)\nonumber \\
 & = & \int dXdx\; B_{m}(x,X,t)\tilde{\rho}_{m}(x,X),\label{barB-map}\end{eqnarray}
 where $\tilde{\rho}_{m}(x,X)=\int dx'f(x,x')\rho_{m}(x',X)$ and
\begin{eqnarray}
 &  & f(x,x')=\frac{1}{(2\pi\hbar)^{N}}\sum_{\lambda\lambda'}\int dzdz'\langle m_{\lambda}|r-\frac{z}{2}\rangle\langle r+\frac{z}{2}|m_{\lambda'}\rangle\nonumber \\
 &  & \qquad\times\langle m_{\lambda'}|r'-\frac{z'}{2}\rangle\langle r'+\frac{z'}{2}|m_{\lambda}\rangle e^{-i(p\cdot z+p'\cdot z')/\hbar}.\end{eqnarray}
 The function $f(x,x')$ can be computed explicitly using Eqs.~(\ref{coordinatereporesentationofmapping})
and (\ref{oscillator}).

\subsection{Evolution equation in mapping basis}

In quantum-classical dynamics the time evolution of an operator may
be described by the quantum-classical Liouville equation,\citep{kapral99}
\begin{eqnarray}
\frac{d}{dt}\hat{B}_{W}(t) & = & \frac{i}{\hbar}[\hat{H}_{W},\hat{B}_{W}(t)]\\
 &  & -\frac{1}{2}(\{\hat{H}_{W},\hat{B}_{W}(t)\}-\{\hat{B}_{W}(t),\hat{H}_{W}\}).\nonumber \end{eqnarray}
 In order to make use of Eq.~(\ref{barB-map}) we must cast this
equation in the mapping basis.

Using the results of the previous subsection, the matrix elements
of any operator $\hat{C}_{W}$ can be written in the form, \begin{equation}
\langle\lambda|\hat{C}_{W}|\lambda'\rangle=\int dqdq'\langle m_{\lambda}|q\rangle\langle q|\hat{C}_{m}|q'\rangle\langle q'|m_{\lambda'}\rangle.\label{Cq}\end{equation}
 Furthermore, if $\hat{C}_{W}=\hat{A}_{W}\hat{B}_{W}$ is a composition
of operators, we have \begin{equation}
\langle\lambda|\hat{C}_{W}|\lambda'\rangle=\langle m_{\lambda}|\hat{C}_{m}|m_{\lambda'}\rangle=\langle m_{\lambda}|\hat{A}_{m}\hat{B}_{m}|m_{\lambda'}\rangle.\label{comp-map}\end{equation}
 Using Eqs.~(\ref{Cq}) and (\ref{comp-map}), we can write the quantum-classical
Liouville equation as \begin{eqnarray}
 &  & \frac{d}{dt}\langle q|\hat{B}_{m}(t)|q'\rangle=\frac{i}{\hbar}\langle q|[\hat{H}_{m},\hat{B}_{m}(t)]|q'\rangle\\
 &  & -\frac{1}{2}(\langle q|\{\hat{H}_{m},\hat{B}_{m}(t)\}|q'\rangle-\langle q|\{\hat{B}_{m}(t),\hat{H}_{m}\}|q'\rangle).\nonumber \end{eqnarray}
 Taking the Wigner transform over the mapping coordinate space we
obtain \begin{eqnarray}
 &  & \frac{d}{dt}B_{m}(x,X,t)=\frac{2}{\hbar}H_{m}\sin(\frac{\hbar\Lambda_{m}}{2})B_{m}(t)\label{eq:pmweq}\\
 &  & \qquad-\frac{\partial H_{m}}{\partial R}\cos(\frac{\hbar\Lambda_{m}}{2})\cdot\frac{\partial B_{m}(t)}{\partial P}+\frac{P}{M}\cdot\frac{\partial B_{m}(t)}{\partial R},\nonumber \end{eqnarray}
 where the negative of the Poisson bracket operator on the mapping
phase space coordinates is defined as $\Lambda_{m}=\overleftarrow{\nabla_{p}}\cdot\overrightarrow{\nabla_{r}}-\overleftarrow{\nabla_{r}}\cdot\overrightarrow{\nabla_{p}}$.
In writing this equation we have used the fact that the Wigner transform
of a product of mapping operators is given by \begin{equation}
(\hat{A}_{m}(X)\hat{B}_{m}(X))_{W}={A}_{m}(x,X)e^{\hbar\Lambda_{m}/2i}{B}_{m}(x,X).\end{equation}
 The Wigner transform of a mapping variable is given by \begin{equation}
A_{m}(x,X)=\frac{1}{2\hbar}\sum_{\lambda\lambda'}A^{\lambda\lambda'}(R)\Big(r_{\lambda}r_{\lambda'}+p_{\lambda}p_{\lambda'}-\hbar\delta_{\lambda\lambda'}\Big),\label{eq:mapA}\end{equation}
 In particular the mapping Hamiltonian takes the form \begin{eqnarray}
H_{m}(x,X) & = & \frac{P^{2}}{2M}+V_{B}(R)\label{eq:mapH}\\
 & + & \frac{1}{2\hbar}\sum_{\lambda\lambda'}h_{\lambda\lambda'}(R)(r_{\lambda}r_{\lambda'}+p_{\lambda}p_{\lambda'}-\hbar\delta_{\lambda\lambda'}),\nonumber \end{eqnarray}
 where $h_{\lambda\lambda'}(R)=\langle\lambda|\hat{p}^{2}/2m+V_{s}(\hat{q})+V_{c}(\hat{q},{R})|\lambda'\rangle$
and we have used the fact that $h_{\lambda\lambda'}=h_{\lambda'\lambda}$.
Given this form of the Hamiltonian one may show that \begin{equation}
H_{m}\Lambda_{m}B_{m}=\frac{1}{\hbar}\sum_{\lambda\lambda'}h_{\lambda\lambda'}(p_{\lambda}\frac{\partial}{\partial r_{\lambda'}}-r_{\lambda}\frac{\partial}{\partial p_{\lambda'}})B_{m},\end{equation}
 \begin{equation}
H_{m}\Lambda_{m}^{2}B_{m}=\frac{1}{\hbar}\sum_{\lambda\lambda'}h_{\lambda\lambda'}(\frac{\partial}{\partial r_{\lambda'}}\frac{\partial}{\partial r_{\lambda}}+\frac{\partial}{\partial p_{\lambda'}}\frac{\partial}{\partial p_{\lambda}})B_{m},\end{equation}
 and \begin{equation}
H_{m}\Lambda_{m}^{n}B_{m}=0.\qquad\ \qquad(\mbox{when}\, n\geq3)\end{equation}
 Then, using these relations, we can simplify Eq.~(\ref{eq:pmweq})
to derive the quantum-classical Liouville equation in the mapping
basis: \begin{eqnarray}
 &  & \frac{d}{dt}B_{m}(x,X,t)=\frac{1}{\hbar}\sum_{\lambda\lambda'}h_{\lambda\lambda'}(p_{\lambda}\frac{\partial}{\partial r_{\lambda'}}-r_{\lambda}\frac{\partial}{\partial p_{\lambda'}})B_{m}(t)\nonumber \\
 &  & \quad+\Big(\frac{P}{M}\cdot\frac{\partial}{\partial R}-\frac{\partial H_{m}}{\partial R}\cdot\frac{\partial}{\partial P}\Big)B_{m}(t)\label{eq:awt2}\\
 &  & \quad+\frac{\hbar}{8}\sum_{\lambda\lambda'}\frac{\partial h_{\lambda\lambda'}}{\partial R}(\frac{\partial}{\partial r_{\lambda'}}\frac{\partial}{\partial r_{\lambda}}+\frac{\partial}{\partial p_{\lambda'}}\frac{\partial}{\partial p_{\lambda}})\cdot\frac{\partial}{\partial P}B_{m}(t).\nonumber \end{eqnarray}
 Since the quantum-classical Liouville equation is exact for an arbitrary
quantum subsystem bilinearly coupled to a harmonic bath, the mapping
version of this equation, Eq.~(\ref{eq:awt2}), is also exact for
such systems.

The first term in Eq.~(\ref{eq:awt2}) is the quantum evolution of
the subsystem in the mapping phase space, while the second term describes
the evolution of the bath where the forces involve the mapping coordinates.
The complicated third term represents the higher-order correlations
between the subsystem and the bath. The evolution equation can be
written more compactly as \begin{eqnarray}
 &  & \frac{d}{dt}B_{m}(x,X,t)=-\{H_{m},B_{m}(t)\}_{x,X}\label{eq:awt3}\\
 &  & \quad+\frac{\hbar}{8}\sum_{\lambda\lambda'}\frac{\partial h_{\lambda\lambda'}}{\partial R}(\frac{\partial}{\partial r_{\lambda'}}\frac{\partial}{\partial r_{\lambda}}+\frac{\partial}{\partial p_{\lambda'}}\frac{\partial}{\partial p_{\lambda}})\cdot\frac{\partial}{\partial P}B_{m}(t)\nonumber \\
 &  & \qquad\qquad\qquad\equiv i{\mathcal{L}}_{m}B_{m}(t),\nonumber \end{eqnarray}
 where $\{A_{m},B_{m}(t)\}_{x,X}$ denotes a Poisson bracket in the
full mapping-bath phase space of the system. The last line of this
equation defines the quantum-classical Liouville operator in the mapping
basis, \begin{equation}
i{\mathcal{L}}_{m}=i{\mathcal{L}}_{m}^{0}+i{\mathcal{L}}_{m}^{\prime},\label{eq:mapL}\end{equation}
 where \begin{eqnarray}
i{\mathcal{L}}_{m}^{0} & = & -\{H_{m},B_{m}\}_{x,X},\quad{\rm and}\\
i{\mathcal{L}}_{m}^{\prime} & = & \frac{\hbar}{8}\sum_{\lambda\lambda'}\frac{\partial h_{\lambda\lambda'}}{\partial R}(\frac{\partial}{\partial r_{\lambda'}}\frac{\partial}{\partial r_{\lambda}}+\frac{\partial}{\partial p_{\lambda'}}\frac{\partial}{\partial p_{\lambda}})\cdot\frac{\partial}{\partial P}.\nonumber \end{eqnarray}
 The complex form of the $i{\mathcal{L}}_{m}^{\prime}$ makes simulation
of the dynamics in the mapping basis difficult. If the evolution equation
is approximated by $i{\mathcal{L}}_{m}^{0}$, simulation of the dynamics
in terms of Newtonian trajectories is straightforward in view of Poisson
bracket form of the resulting equation of motion. The validity of
this approximation must be determined for specific applications. In
the next section we apply this equation to the spin-boson model and
show that accurate results can be obtained when the last term on the
right side of this equation is neglected.

\section{Spin-boson model}

The spin-boson model is often used as a test case for quantum simulations
of many-body systems and we present the results of simulations of
this model using the quantum-classical Liouville equation in the mapping
basis. The spin-boson model describes a two-level system bilinearly
coupled to a harmonic bath of $N_{B}$ oscillators with masses $M_{j}$
and frequencies $\omega_{j}.$ The system Hamiltonian is given by
\begin{equation}
{\bf H}_{W}=\left(\begin{array}{cc}
{H}_{B}+\hbar\gamma(R) & -\hbar\Omega\\
-\hbar\Omega & {H}_{B}-\hbar\gamma(R)\end{array}\right),\label{eq:tlsH}\end{equation}
 where ${H}_{B}=\sum_{j}\left({P}_{j}^{2}/2M_{j}+M_{j}\omega_{j}^{2}R_{j}^{2}/2\right)$
and $\gamma({R})=-\sum c_{j}R_{j}$. The energy gap of the isolated
two-state system is $2\hbar\Omega$. From Eqs.~(\ref{eq:mapH}) and
(\ref{eq:tlsH}) we can obtain \begin{equation}
H_{m}=H_{B}+\frac{1}{2}\gamma(R)(r_{1}^{2}+p_{1}^{2}-r_{2}^{2}-p_{2}^{2})-\Omega(r_{1}r_{2}+p_{1}p_{2}).\end{equation}
 Equation~(\ref{eq:awt3}) is an exact evolution equation for the
spin-boson model. Previous simulations of the quantum-classical Liouville
equation in the adiabatic basis have been carried out using a Trotter-based
scheme and were able to reproduce the exact results for a wide range
of system parameters.\citep{mackernan08} Consequently, the results
presented here can be viewed as a test of the utility of the simulation
schemes that use the mapping basis to represent quantum-classical
Liouville dynamics.

As in previous studies we assume that the initial density matrix is
uncorrelated so that the subsystem is in the ground state and bath
is in thermal equilibrium, namely, \begin{equation}
{\mbox{\boldmath{$\rho$}}}_{W}(0)={\mbox{\boldmath{$\rho$}}}_{s}(0)\rho_{B}(X),\qquad{\mbox{\boldmath{$\rho$}}}_{s}(0)=\left(\begin{array}{cc}
1 & 0\\
0 & 0\end{array}\right),\end{equation}
 where the Wigner distribution of the bath $\rho_{B}(X)$ is given
by\citep{imre67,kim06} \begin{equation}
\rho_{B}(X)=\prod_{j=1}^{N}\frac{\beta\omega_{j}}{2\pi u_{j}''}\exp[-\frac{\beta}{u_{j}''}\{\frac{1}{2M_{j}}P_{j}^{2}+\frac{1}{2}M_{j}\omega_{j}^{2}R_{j}^{2}\}],\end{equation}
 with $u_{j}''=u_{j}\coth u_{j}$ and $u_{j}=\beta\hbar\omega_{j}/2$.
The subsystem initial density matrix in the Wigner-transformed mapping
basis is \begin{equation}
\rho_{sm}(x)=(2\pi\hbar)^{-2}(2\hbar)^{-1}(r_{1}^{2}+p_{1}^{2}-\hbar).\end{equation}
 Using the results in Eq.~(\ref{barB-map}), the time evolution of
the population difference between the ground and excited states, given
by the expectation value of Pauli matrix $\hat{\sigma}_{z}$, can
be written as \begin{eqnarray}
\overline{\sigma_{z}(t)} & = & \sum_{\lambda\lambda'}\int dX\sigma_{z}^{\lambda\lambda'}(t)\rho_{s}^{\lambda'\lambda}(0)\rho_{B}(X)\\
 & = & \int dXdx\;\sigma_{zm}(x,X,t)\tilde{\rho}_{sm}(x)\rho_{B}(X),\nonumber \end{eqnarray}
 where $\tilde{\rho}_{sm}(x)$ has the explicit form \begin{equation}
\tilde{\rho}_{sm}(x)=\frac{2}{\hbar^{3}\pi^{2}}(r_{1}^{2}+p_{1}^{2}-\frac{\hbar}{2})e^{-(r^{2}+p^{2})/\hbar}.\end{equation}
 The initial value of the Wigner-transformed mapping representation
of $\hat{\sigma}_{z}$ is $\sigma_{zm}(x)=(2\hbar)^{-1}(r_{1}^{2}+p_{1}^{2}-r_{2}^{2}-p_{2}^{2})$.

\begin{figure}
\includegraphics[clip,width=0.8\columnwidth]{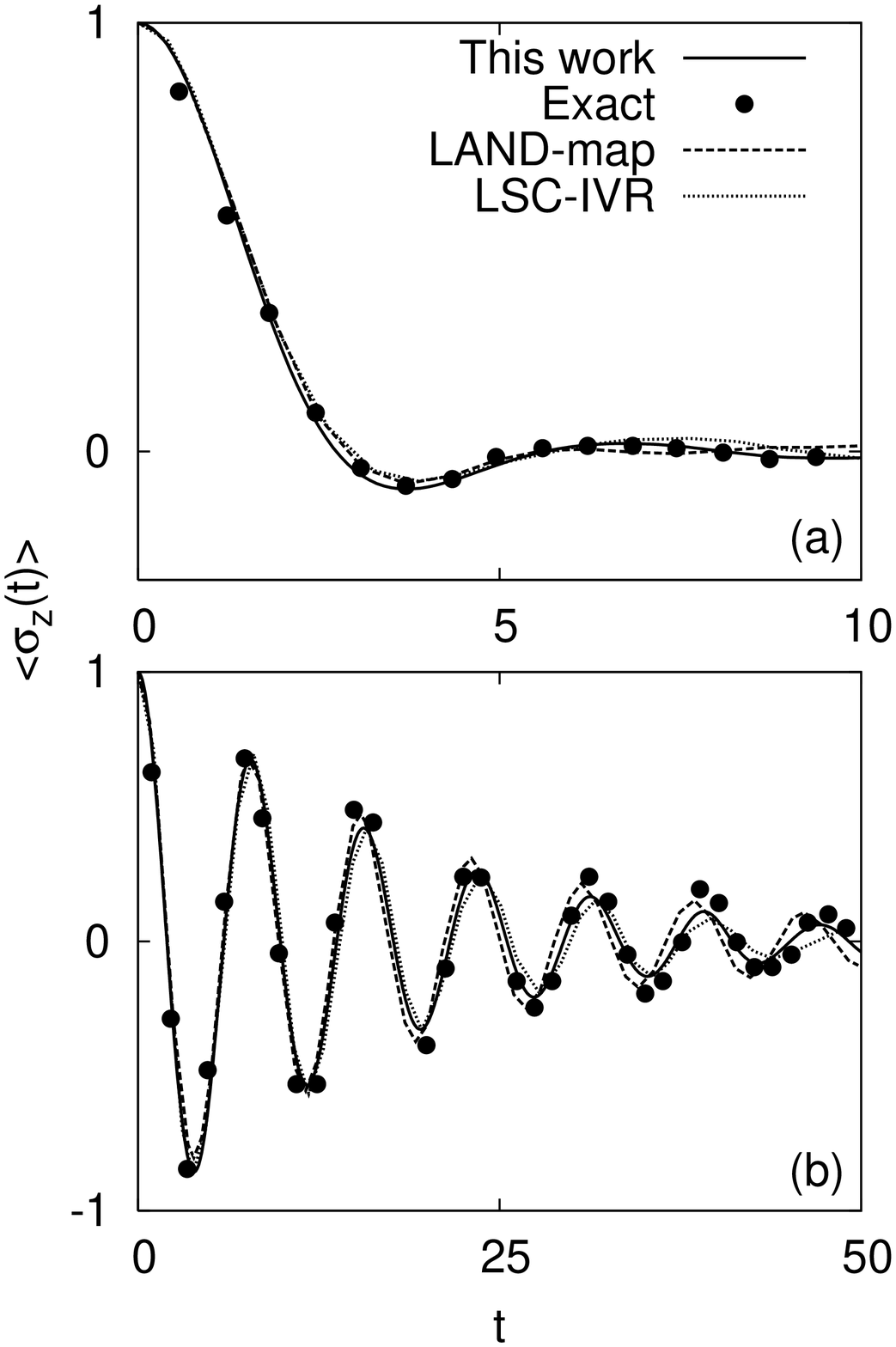}

\caption{Electronic population difference $\langle\sigma_{z}(t)\rangle$ as
a function of $t$ for two dimensionless parameter sets: $\Omega=0.4$,
$\xi=0.09$, and $\beta=0.25$ (a) or $12.5$ (b). The solid points
are exact results\citep{makarov94}, the dashed lines are the LAND-map
results\citep{bonella05} and the dotted lines are the LSC-IVR results\citep{sun98}.
\label{fig:1}}

\end{figure}

To compute $\overline{\sigma_{z}(t)}$ we need to solve for $\sigma_{zm}(x,X,t)$
using Eq.~(\ref{eq:awt3}). This equation is difficult to solve because
of the structure of the last term of the quantum-classical Liouville
operator in the mapping basis. For the spin-boson model one may show
by direct calculation for short times that the last term does not
contribute until the fifth-order initial derivative of $\sigma_{zm}(x)$.
This suggests that it may be possible to obtain a useful approximate
solution by neglecting the last term in the evolution equation~(\ref{eq:awt3})
so that \begin{equation}
\frac{d}{dt}\sigma_{zm}(x,X,t)\approx i{\mathcal{L}}_{m}^{0}\sigma_{zm}(t).\end{equation}
 The dynamical variable $\sigma_{zm}(x,X,t)$ evolves by Newtonian
equations of motion and admits a solution in terms of characteristics.
The corresponding set of ordinary differential equations is \begin{eqnarray}
\frac{dr_{\lambda}(t)}{dt} & = & \frac{1}{\hbar}\sum_{\lambda'}h_{\lambda\lambda'}(R(t))p_{\lambda'}(t),\nonumber \\
\frac{dp_{\lambda}(t)}{dt} & = & -\frac{1}{\hbar}\sum_{\lambda'}h_{\lambda\lambda'}(R(t))r_{\lambda'}(t),\nonumber \\
\frac{dR(t)}{dt} & = & \frac{P(t)}{M},\quad\frac{dP(t)}{dt}=-\frac{\partial H_{m}}{\partial R(t)}.\end{eqnarray}

Using this result, we obtain the simple form for the expectation value,
\begin{eqnarray}
 &  & \overline{\sigma_{z}(t)}=\left(\frac{1}{\pi^{2}\hbar^{4}}\right)\int dxdX\;\rho_{B}(X)e^{-(r^{2}+p^{2})/\hbar}\label{eq:integ1}\\
 &  & \quad\times(r_{1}^{2}+p_{1}^{2}-\frac{\hbar}{2})(r_{1}(t)^{2}+p_{1}(t)^{2}-r_{2}(t)^{2}-p_{2}(t)^{2}).\nonumber \end{eqnarray}

The linear coupling in the spin-boson model is characterized by an
Ohmic spectral density, $J(\omega)=\pi\sum c_{j}^{2}/(2M_{j}\omega_{j})\delta(\omega-\omega_{j})$,
where $c_{j}=(\xi\hbar\Delta\omega M_{j})^{1/2}\omega_{j}$, $\omega_{j}=-\omega_{c}\ln\left(1-j\Delta\omega/\omega_{c}\right)$
and $\Delta\omega=\omega_{c}\left(1-e^{-\omega_{max}/\omega_{c}}\right)/N_{B}$
with $\omega_{c}$ the cut-off frequency and $\xi$ the Kondo parameter.
\citep{makri99} We used $N_{B}=20$ and $\omega_{max}=4\omega_{c}$.
Dimensionless units with time scaled by $\omega_{c}$ are used in
the calculations below. The equations of motion were integrated using
the velocity Verlet algorithm with time step $\Delta t=0.1$.

\begin{figure}[htbp]
\includegraphics[clip,width=0.8\columnwidth]{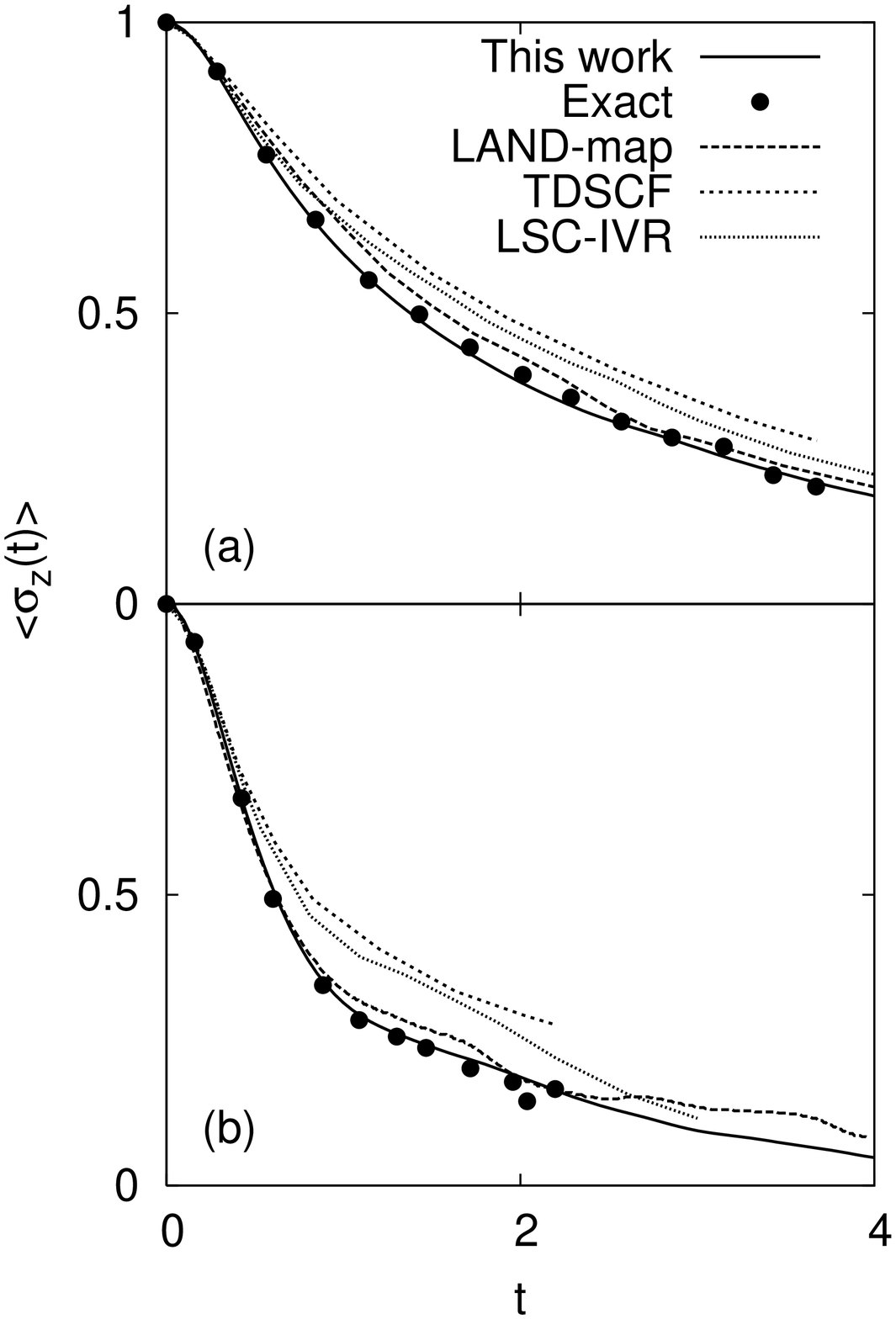}

\caption{Electronic population difference $\langle\sigma_{z}(t)\rangle$ as
a function of $t$ for two parameter sets: $\xi=2$, $\beta=0.25$,
and $\Omega=0.8$ (a) or $1.2$ (b). The solid points are exact results\citep{egger94},
the dashed lines are the LAND-map results\citep{bonella05}, the dot-dashed
lines are the TDSCF results\citep{stock95} and the dotted lines are
the LSC-IVR results\citep{sun98}. \label{fig:2}}

\end{figure}

The expectation value $\overline{\sigma_{z}}(t)$ in Eq.~(\ref{eq:integ1})
may be computed by sampling initial bath and mapping variables from
Gaussian distributions, reweighting to account for the form of the
initial density matrix, and computing $\sigma_{zm}(x(t))$. We have
also carried out the calculations using focused initial conditions\citep{bonella03,bonella05}
where the state mapping variables are initially taken to be $r_{1}=1$,
$p_{1}=1$, $r_{2}=0$ and $p_{2}=0$ when state 1 is initially occupied.
Both sampling methods yield comparable results but focused initial
conditions require about a factor of ten fewer trajectories to obtain
converged results for this model.

\begin{figure}[htbp]
\includegraphics[clip,width=0.8\columnwidth]{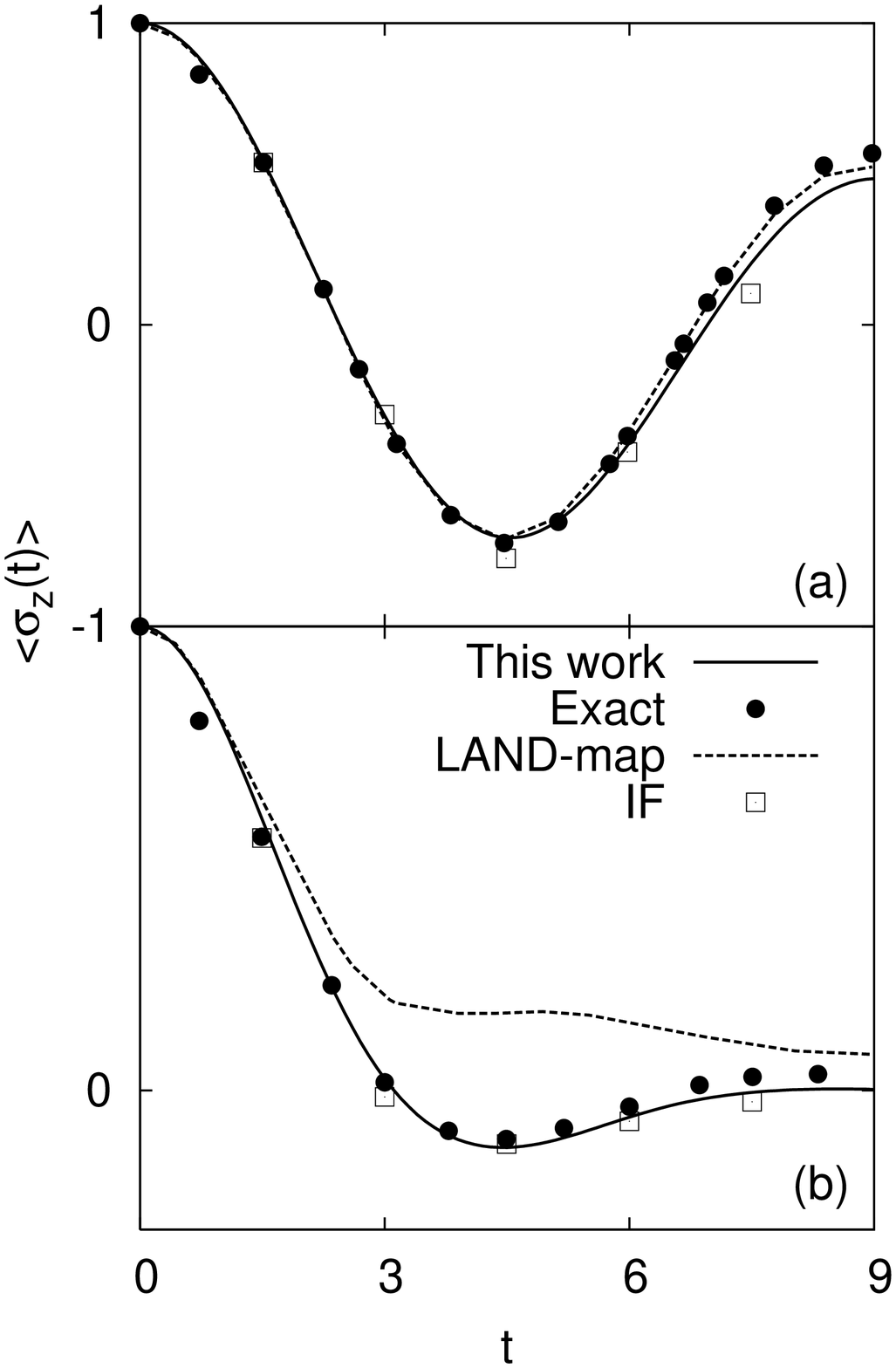}

\caption{Electronic population difference $\langle\sigma_{z}(t)\rangle$ as
a function of $t$ for two parameter sets: $\Omega=1/3$, $\beta=3$,
and $\xi=0.1$ (a) or $0.5$ (b). The solid points are exact results\citep{thompson99},
the dashed lines are the LAND-map results\citep{bonella05}, and the
open squares are the results obtained using imaginary time path integral
semiclassical influence functional formalism with four time slices\citep{thompson99}.
\label{fig:3}}

\end{figure}

We tested our method for the parameters for which numerically exact
results are available. Approximately $10^{4}$ trajectories were used
to obtain the results in the figures. Comparable results can be obtained
even with ten times fewer trajectories. In Fig.~\ref{fig:1} the
results are compared for weak system-bath coupling with $\xi=0.09$.
The adiabatic energy gap is chosen as $\Omega=0.4$. For high temperatures,
the time-dependent population difference exhibits incoherent behavior
as in Fig.~\ref{fig:1}(a). Our results, as well as those of other
methods such as LAND-map and LSC-IVR, show excellent agreement with
the numerically exact results\citep{makarov94} for the high-temperature,
weak-coupling case. The reproduction of the coherent or oscillatory
behavior at low temperatures shown in Fig.~\ref{fig:1}(b) is a more
severe test, especially at long times. Our results predict the correct
frequency of oscillations but the magnitude of the oscillations are
somewhat smaller at long times.

In Fig.~\ref{fig:2}, we plot $\langle\sigma_{z}(t)\rangle$ for
a rather high friction constant, $\xi=2$, at high temperature of
$\beta=0.25$. One can see that the accuracy of our results does not
change for strong system-bath coupling and is consistently better
than other approaches.

As a final test, in Fig.~\ref{fig:3} we show $\langle\sigma_{z}(t)\rangle$
for two friction constants, $\xi=0.1$ and 0.5, for a relatively low
temperature, $\beta=3$. The LAND-map approach predicts the slow incoherent
decay instead of oscillation around zero. This discrepancy was attributed
to the linearization approximation which underestimates the coherent
dynamics. Our results again show reliable accuracy both for weak and
strong coupling. Our results are compared with those using the semiclassical
influence functional formalism with four time slices. Similar accuracy
is obtained.

\section{Conclusion}

The representation of the quantum-classical Liouville equation in
the mapping Hamiltonian basis provides another way to simulate nonadiabatic
dynamics. The complicated form of $i{\mathcal{L}}_{m}^{\prime}$ in
Eq.~(\ref{eq:mapL}) in this basis leads to difficulties in the construction
of simulation algorithms. If this term is neglected and the evolution
is approximated by $i{\mathcal{L}}_{m}^{0}$ the dynamics may be computed
easily using an ensemble of trajectories. This is an excellent approximation
for the spin-boson model and leads to a simulation scheme for nonadiabatic
dynamics that does not involve surface-hopping. The extent to which
this approximation is applicable to more general systems remains to
be determined. The work also suggests that it may be possible to construct
simulation algorithms that use evolution under $i{\mathcal{L}}_{m}^{0}$
as a zeroth order scheme about which corrections can be computed.
The utility of the mapping formulation of the quantum-classical Liouville
equation for the computation of general correlation functions is also
another topic that is worth pursuing.

\begin{acknowledgments}
This work was supported in part by a grant from the Natural Sciences
and Engineering Research Council of Canada. We would like to thank
Sara Bonella for many useful discussions.
\end{acknowledgments}
\bibliographystyle{apsrev}
\bibliography{totalrefs}

\newpage{}

\listoffigures
\end{document}